\documentclass[%
reprint,
superscriptaddress,
showpacs,preprintnumbers,
 amsmath,amssymb,
 aps,
prl,
floatfix,
]{revtex4-1}

\usepackage{graphicx}
\usepackage{dcolumn}
\usepackage{bm}
\usepackage{hyperref}

\begin{document}


\title{Voltage-Controllable Colossal Magnetocrystalline Anisotropy in Single Layer Transition Metal Dichalcogenides} 


\author{Xuelei Sui}
\affiliation{Department of Physics and State Key Laboratory of Low-Dimensional Quantum Physics, Tsinghua University, Beijing 100084, People's Republic of China}
\affiliation{Computational Science Research Center, Beijing 100084, People's Republic of China}
\author{Tao Hu}
\affiliation{Computational Science Research Center, Beijing 100084, People's Republic of China}
\affiliation{Department of Chemistry and Biochemistry, California State University Northridge, California, Los Angeles 18111, United States}
\author{Jianfeng Wang}
\affiliation{Department of Physics and State Key Laboratory of Low-Dimensional Quantum Physics, Tsinghua University, Beijing 100084, People's Republic of China}
\author{Bing-Lin Gu}
\affiliation{Department of Physics and State Key Laboratory of Low-Dimensional Quantum Physics, Tsinghua University, Beijing 100084, People's Republic of China}
\affiliation{Institute for Advanced Study, Tsinghua University, Beijing 100084, People's Republic of China}
\affiliation{Collaborative Innovation Center of Quantum Matter, Beijing 100084, People's Republic of China}
\author{Wenhui Duan}
\email{dwh@phys.tsinghua.edu.cn}
\affiliation{Department of Physics and State Key Laboratory of Low-Dimensional Quantum Physics, Tsinghua University, Beijing 100084, People's Republic of China}
\affiliation{Institute for Advanced Study, Tsinghua University, Beijing 100084, People's Republic of China}
\affiliation{Collaborative Innovation Center of Quantum Matter, Beijing 100084, People's Republic of China}
\author{Mao-sheng Miao}
\email{mmiao@csun.edu}
\affiliation{Department of Chemistry and Biochemistry, California State University Northridge, California, Los Angeles 18111, United States}
\affiliation{Computational Science Research Center, Beijing 100084, People's Republic of China}
\date{\today}

\begin{abstract}
Materials with large magnetocrystalline anisotropy and strong electric field effects are highly needed to develop new types of memory devices based on electric field control of spin orientations. Instead of using modified transition metal films, we propose that certain monolayer transition metal dichalcogenides are the ideal candidate materials for this purpose. Using density functional calculations, we show that they exhibit not only a large magnetocrystalline anisotropy (MCA), but also colossal voltage modulation under external field. Notably, in some materials like CrSe$_{2}$ and FeSe$_{2}$, where spins show a strong preference for in-plane orientation, they can be switched to out-of-plane direction. This effect is attributed to the large band character alteration that the transition metal $d$-states undergo around the Fermi energy due to the electric field. We further demonstrate that strain can also greatly change MCA, and can help to improve the modulation efficiency while combined with an electric field.
\end{abstract}


\maketitle

Enormous efforts, both experimental and theoretical, have been spent to improve the efficiency of magnetization control in nanoscale systems\cite{nature,mutiferroics}. Conventional techniques like magnetic-field-induced magnetization switch and spin-current induced torque in magnetic tunnel junctions\cite{spin}, both have a complex design and are very power-consuming. In comparison, controlling the spin orientation by applying electric fields to materials with large magnetocrystalline anisotropy (MCA) is a new and promising approach. This method has the advantage of ultra-low power consumption and strong coherence of the individual spins. Recently, this has been demonstrated experimentally in itinerant magnetic FePt and FePd ultrathin films with liquid interfaces\cite{science}. Soon after, electric-field-controlled MCAs were also reported for few-monolayers-thick magnetic metals\cite{surface_metal,monolayer,fewlayeriron} and alloys\cite{FeCo,Au/FeCo}, nano-junctions\cite{junction,junction2}, defected graphene\cite{Dimer}, and thin films\cite{Kioussis}. In most of these materials, the magnetism comes from the $3d$ transition metals and the MCA arises from the strong spin-orbit coupling (SOC) due to the alloying with heavy elements.  Markedly, the MCA in these materials is still low, and the spin states are vulnerable to thermal fluctuations. The electric field effect also needs large improvements for these materials to effectively manipulate the spin orientation. Furthermore, other problems persist in thin metal films and surfaces, including the difficulty of growing high quality samples, the high reactivity while exposed to air and liquid, and especially the strong screening to the applied electric field.

In contrast to ultra-thin metal films, many two-dimensional (2D) materials can be more easily fabricated in large quantity and high quality\cite{geim2007rise,butler2013progress,two-dimensional}. Many of them are quite stable and their screening effect to electric field is relatively small. 2D materials, such as graphene, boron-nitride and transition metal dichalcogenides (TMDs) have shown high stability and superior transport properties\cite{yan2007,defects}. Furthermore, mechanical exfoliation or chemical synthesis can be used to produce monolayer TMDs flakes of high purity, such as MoS$_2$\cite{synthetic}, MoSe$_2$, TaS$_2$, TaSe$_2$, and NiTe$_2$\cite{exfoliation,mechanical,single_layer}. Recently, a thorough computational study was carried out for more than 30 monolayer TMDs with varying combinations of transition metal and chalcogen atoms (S, Se, or Te)\cite{stable}. Depending on $d$ band filling, TMD monolayers can be magnetic. The examples also include the well studied VSe$_2$, TaS$_2$ and TaSe$_2$\cite{VS2,grain,magnetic}. A couple of recent studies used TMDs as the supporting materials for the MCA centers\cite{Fe/MoS,Fe/TaS}. Since many of the magnetic TMDs contain heavy elements such as Se and Te, which indicates strong SOC, we propose these monolayer materials may possess large voltage-controllable MCA.

In this letter, we conduct a systematic first-principles study of the MCA of single-layer TMDs and related materials with and without external electric field. The selected materials include AX$_2$ (A = Sc, Ti, V, Cr, Mn, Fe, Co, Ni, X = Se, Te), TaS$_2$, TaSe$_2$ and also FeI$_2$. The results clearly demonstrate a large MCA for a number of AX$_2$ monolayers, including CrSe$_2$, FeSe$_2$, FeTe$_2$, TaS$_2$, TaSe$_2$, and FeI$_2$. Especially, the MCA for some AX$_2$ shows a very strong electric field dependence. For CrSe$_2$, FeSe$_2$, FeTe$_2$, and FeI$_2$, the electric field can change the sign of MCA, which can be used to switch preferred spin orientation. Our study illustrates the promising potential of electric-field-switching of magnetization orientation in these 2D materials. Furthermore, we demonstrate that the strain effect on MCA is also strong.

\begin{figure} [ht]
\centering
 \includegraphics[width=0.35\textwidth]{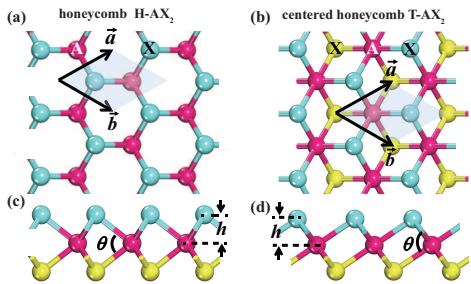}\\
 \caption{(Color online) Atomic structures of single layer AX$_2$. (a) and (c) are the top and side views of H phase, (b) and (d) are top and side views of T phase. The shaded areas show the primitive unit cells with Bravais lattice vectors $\vec{a}$ and $\vec{b}$ ($|\vec{a}| = |\vec{b}|$). $h$ is the vertical distance between the chalcogen atom and transition metal atom and $\theta$ is the X-A-X bond angle across the top and the bottom layers. Magenta spheres represent transition-metal atoms (A) while blue and yellow spheres represent the top and the bottom layer chalcogen atoms (X), respectively.}
 \label{fig:Figure1}
\end{figure}

All the calculations are performed within the framework of density functional theory (DFT)\cite{DFT} as implemented in the Vienna \emph{ab initio} simulation package (VASP)\cite{VASP}. The PAW\cite{PAW} potentials are used to describe the ionic potential of all the atoms. We employed the Perdew-Burke-Ernzerhof (PBE)\cite{PBE} generalized gradient approximation (GGA) for the exchange correlation functional. In order to treat the strong on-site Coulomb interaction of $3d$ metals, we employed the GGA + U\cite{LDA+U} method. The $U$ values are tested for all the compounds by comparing the resulted magnetic moment with those obtained by Heyd-Scuseria-Ernzerhof (HSE)\cite{HSE} hybrid functional. For most of the cases, we find that $U$ = 2 eV yield magnetic moments in good comparison with HSE.

Several factors are important for the selection of candidate 2D TMD systems. 1) The material should be thermodynamically stable or metastable with a low energy. Ideally, the TMD could exist naturally in single layer or in layered bulk structure that can be exfoliated into single layer. 2) The material should be spin polarized, therefore the TMD materials containing 3d transition metals especially V, Cr, Mn, Fe, Co are particullarly interesting. 3) The material should have large SOC, therefore should contain heavy elements. Our work will then focus on late chalcogens such as Se and Te. Some compounds, although not dichalcogenides, adopt the same layered structure as TMD and satisfy the above criteria. One such example is FeI$_2$, in which Iodine atoms occupy the same lattice sites as chacogens. We therefore also include these layered structures in our study.

We first investigate the stability of single layer AX$_2$ by comparing their energies with existing A$_n$X$_m$ compounds. For an existing A$_m$X$_n$ compound, we assume a reaction between A$_m$X$_n$ and elemental solid A, $2/n$ A$_m$X$_n$ + $[(n-2m)/n]$ A $\rightarrow$ AX$_2$. The formation energy $E_f$ is defined as: $E_f$ = $E_{\text {AX}_2}$ - $2/n$ $E_{\text A_m\text X_n}$ - $[(n-2m)/n]$ $E_\text A$, where $E_{\text A_m\text X_n}$ and $E_\text A$ are the total energies of A$_m$X$_n$ compound and solid A, respectively. For 3D AX$_2$ compounds, $E_f$ is actually the energy difference between single layer and bulk AX$_2$. There are two different types of single-layer structures for AX$_{2}$ compounds: a honeycomb H structure with point-group symmetry of D$_{3h}$ (trigonal prismatic coordination, Fig. 1(a) and (c)) and a centered honeycomb T structure with D$_{3d}$ symmetry (octahedral coordination, Fig. 1(b) and (d)). Both structures are included in our study.

The calculated results are shown in Table 1. We would like to point out that several AX$_2$ compounds are already known and exhibit layered structure, including CrSe$_2$, TaS$_2$, TaSe$_2$, and FeI$_2$. Our calculations show that the energy difference between single layer and bulk structure is very small, indicating that CrSe$_2$ monolayer can be fabricated by mechanical exfoliation. Single layer TaS$_2$ and TaSe$_2$ have been fabricated and their properties have been studied\cite{thin_TaS2}. Furthermore, we also examine the dynamic stability of these compounds by calculating the phonon spectra of optimized TMDs in both H and T configurations (see Supporting information Fig. S1). Our results reveal that among the 36 structures considered, 31 of them are dynamically stable. Depending on the coordination and oxidation state of the transition metal atoms, TMDs can be either metallic or semiconducting.

\begin{table}[tbp]
  \caption{Properties of magnetic AX$_2$ monolayers, including formation energies ($E_f$), magnetic energies ($E_{\text{FM}}$ and $E_{\text{AFM}}$) and magnetic moments ($M$) for free-standing single layer AX$_2$ in H and T phases\textsuperscript{\emph{a}}}
  \label{tb:tb1}
\begin{tabular}{cccccccc}
\hline
System &$E_f$ &$E_{\text{FM}}$ &$E_{\text{AFM}}$ &$M_{\text{GGA}}$ &$M_{U=2}$ &$M_{\text{HSE}}$ &\\
\hline
H-ScSe$_2$  &2.29\textsuperscript{\emph{c}}   &-0.32 &-0.16 &1.00  &1.00 &1.00   &S\\
H-ScTe$_2$  &1.85\textsuperscript{\emph{c}}   &-0.25 &-0.14  &1.00  &1.00 &1.00  &S\\
H-VSe$_2$   &-0.01\textsuperscript{\emph{b}}   &-0.36 &-0.15 &1.00  &1.00 &1.00  &S\\
T-VSe$_2$   &-0.02\textsuperscript{\emph{b}}   &-0.44 &-0.29  &0.61  &1.15 &1.26  &M\\
H-VTe$_2$   &0.12\textsuperscript{\emph{b}}   &-0.41 &-0.26 &1.00  &1.00 &1.00  &S\\
T-VTe$_2$   &-0.04\textsuperscript{\emph{b}}   &-0.66 &-0.21  &0.92  &1.50 &1.52  &M\\
T-CrSe$_2$  &0.00\textsuperscript{\emph{b}}   &-1.79 &-0.09  &2.16  &2.42 &2.69  &M\\
T-CrTe$_2$  &0.08\textsuperscript{\emph{b}}   &-1.91 &-1.88 &2.43  &2.66 &2.69  &M\\
T-MnSe$_2$  &-0.33\textsuperscript{\emph{b}}   &-2.38 &-0.19 &2.81  &3.00 &3.00  &H\\
H-MnTe$_2$  &0.13\textsuperscript{\emph{b}}   &-2.39 &-0.06 &2.57  &3.00 &3.00  &H\\
T-MnTe$_2$  &-0.10\textsuperscript{\emph{b}}  &-2.76 &-0.44 &2.82  &3.17 &3.10  &M\\
H-FeSe$_2$  &0.26\textsuperscript{\emph{b}}   &-0.92 &-0.45 &1.99  &2.00 &2.00  &H\\
H-FeTe$_2$  &0.20\textsuperscript{\emph{b}}   &-1.04 &-0.55 &1.80  &2.00 &2.00  &H\\
H-TaS$_2$   &0.01\textsuperscript{\emph{b}}   &-0.06 &-0.02 &0.02  &0.75 &0.87  &M\\
H-TaSe$_2$  &0.01\textsuperscript{\emph{b}}   &-0.08 &-0.02 &0.00  &0.83 &1.00  &M\\
T-FeI$_2$   &0.03\textsuperscript{\emph{b}}   &-1.79 &1.20  &4.00  &4.00 &4.00  &S\\
\hline
\end{tabular}\\
\begin{flushleft}
\footnotesize\noindent\textsuperscript{\emph{a}} $E_f$ (in units of eV) is the formation energy of single layer AX$_2$. \textsuperscript{\emph{b}} AX$_2$ compound exists, \textsuperscript{\emph{c}} AX$_2$ compound does not exist. $E_{\text{FM}}$, $E_{\text{AFM}}$ (in units of eV) are the relative energies of ferromagnetic and anti-ferromagnetic state with respect to the non-magnetic state. $M_{\text{GGA}}$, $M_{U = 2}$ and $M_{\text{HSE}}$ (in units of $\mu_B$) are the values of magnetic moment calculated by GGA, GGA + U and HSE functionals, respectively. S, M and H in the last column stand for semiconducting, metallic and half metallic, respectively.
\end{flushleft}
\end{table}

Now we will focus on the magnetic AX$_2$ monolayers. Table 1 gives the calculated magnetic moments for all magnetic AX$_2$ we have considered. The results obtained by GGA~+~U method ($U$ = 2 eV) compare very well with the HSE results. Magnetic moments of MnX$_2$ and FeX$_2$ (CrX$_2$) are around 3 $\mu_B$ and 2 $\mu_B$, respectively, while others are about 1 $\mu_B$. The magnetic moment is an integer number if the material is semiconducting or half metallic. In latter case, one spin channel is metallic while the other one has a gap (see Fig. S2 in SI). In order to examine the magnetic ordering, we consider two spin configurations of ferromagnetic (FM) and antiferromagnetic (AFM). Both of the energies referring to nonmagnetic state are listed.  By comparing the energies of the FM and AFM states, we find that all magnetic 2D AX$_{2}$ are in FM state.

\begin{figure} [ht]
 \centering
 \includegraphics[width=0.28\textwidth]{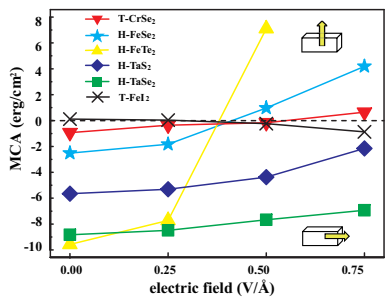}\\
 \caption{(Color online) The calculated MCA (in units of erg/cm$^2$) as a function of electric field ($F=$ 0, 0.25, 0.5 and 0.75 V/{\AA}) for T-FeI$_2$ (black lines), T-CrSe$_2$ (red lines), H-FeSe$_2$ (cyan lines), H-FeTe$_2$ (yellow lines), H-TaS$_2$ (blue lines) and H-TaSe$_2$ (green lines) (H-FeTe$_2$ is unstable when $F =$ 0.75 V/{\AA}). Inset: Schematic diagrams for two magnetic orientations, the upper and lower ones represent the magnetization aligned along the out-of-plane [001] and in-plane [100] directions, respectively.}
 \label{fig:Figure2}
\end{figure}

For the next step, we examine the MCA and its dependence with the electric field for all magnetic AX$_{2}$ monolayers. For a surface of area $S$ (in units of cm$^2$), magnetocrystalline anisotropy is the total energy difference between two magnetic states where the magnetization is aligned along the [100] or [001] direction, i.e., MCA $=( E_{[100]} - E_{[001]})/S$. The results for a number of selected compounds are presented in Fig. 2 (please see Supporting Information Table S1 for MCA values for all studied magnetic AX$_2$ under zero and finite electric fields). As shown in the figure, many 2D AX$_2$ compounds show very large MCAs at zero field. Especially for H-FeTe$_{2}$ and H-TaSe$_{2}$, their MCAs are as large as $-9.58$ and $-8.84$ erg/cm$^{2}$, respectively. Negative values mean that spins favor the in-plane ($x$ or $y$) orientations. These values are one order of magnitude larger than those of transition metal thin films. For example, a Pd{-capped, nine monolayers thick FePd film exhibits an MCA value of 0.86 erg/cm$^{2}$\cite{Kioussis}. A gold (Au) capped FeCo film has a MCA value of -0.56 erg/cm$^{2}$\cite{Au/FeCo}. It is well known that large MCA is originated from the strong coupling between the local spin (magnetic moments) and the orbital moments. Inclusion of heavy elements in the system, like Pd or Au capping in the metal films, can greatly enhance MCA. However, as shown in our current work, forming direct chemical bonds with heavy elements such as Se or Te in AX$_2$ is a more effective way to improve MCA, likely due to the stronger coupling between spin and orbital moments.

As shown in Fig.~\ref{fig:Figure2}, 2D AX$_{2}$ compounds show exceptionally large voltage modulation with their magnetocrystalline anisotropy. The MCA values of all five compounds shown in the figure vary dramatically with the increasing external field. For example, the MCA of H-TaS$_{2}$ changes from $-5.65$ erg/cm$^{2}$ to $-2.16$ erg/cm$^{2}$ while electric field increases from 0 to 0.75 V/{\AA}. More interestingly, MCA values increase monotonically and change sign from negative to positive for T-CrSe$_{2}$ (from $-0.93$ erg/cm$^{2}$ to 0.65 erg/cm$^{2}$), H-FeSe$_{2}$ (from $-2.52$ erg/cm$^{2}$ to 4.19 erg/cm$^{2}$) and H-FeTe$_{2}$ (from $-9.58$ erg/cm$^{2}$ to 7.11 erg/cm$^{2}$) under finite external field. These results clearly show that we can effectively switch the magnetization orientation from in-plane [100] to out-of-plane [001] direction in some 2D AX$_2$ materials by applying electric field. In comparison, an earlier work showed that MCA of monolayer Fe (001) changes from 0.39 erg/cm$^{2}$ (0.2 meV/atom) to $-0.19$ erg/cm$^{2}$ ($-0.1$ meV/atom) while electric field varies from 0 to 1.35 V/{\AA}\cite{monolayer}. 
\begin{figure} [ht]
 \centering
 \includegraphics[width=0.40\textwidth]{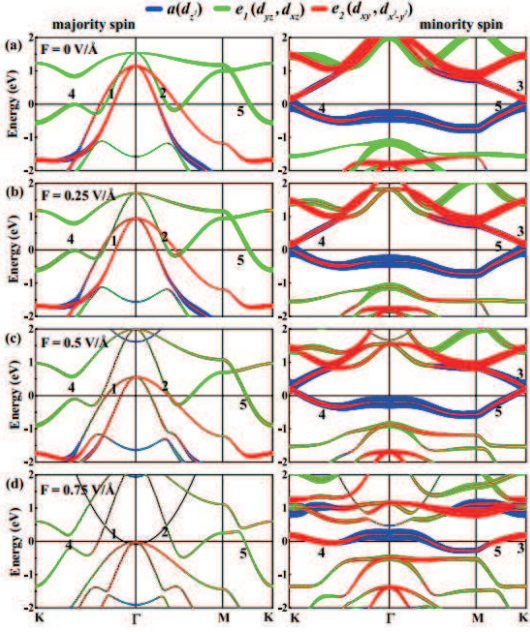}\\
 \caption{(Color online) Energy- and $k$-resolved distribution of orbital character for electronic structures along high symmetry lines for Fe $d$-orbitals in FeSe$_2$ compound under electric field of (a) 0, (b) 0.25, (c) 0.5 and (d) 0.75 V/{\AA}, respectively. The left and right subpanels are majority and minority spin bands, respectively. The thickness represents the amplitude of the orbital character. Numerals refer to points on high-symmetry lines (HSL$n$, $n=1$-5) where there are large changes in MCA.}
 \label{fig:Figure3}
\end{figure}

Now we will discuss the possible microscopic mechanism of electric-field-driven modulation of MCA. It is commonly known that the modification of MCA is caused by the changes in the relative occupation of transition metal $d$ orbitals\cite{fewlayeriron}. Assuming the SOC is a perturbation term of the Hamiltonian, MCA can be expressed by the coupling terms between the occupied and unoccupied states through the orbital angular momentum operators $\hat{L}_{z}$ and $\hat{L}_{x}$ and on the energy difference between these states\cite{MCA,origin}, namely
\begin{eqnarray}
\text{MCA}= {{\xi }^{2}}\sum\limits_{\rm o,u}{\frac{{{\left| \left\langle  \Psi _{\rm o}^{\uparrow \left( \downarrow  \right)} \right|{{{\hat{L}}}_{z}}\left| \Psi _{\rm u}^{\uparrow \left( \downarrow  \right)} \right\rangle  \right|}^{2}}-{{\left| \left\langle  \Psi _{\rm o}^{\uparrow \left( \downarrow  \right)} \right|{{{\hat{L}}}_{x}}\left| \Psi _{\rm u}^{\uparrow \left( \downarrow  \right)} \right\rangle  \right|}^{2}}}{E_{\rm u}^{\uparrow \left( \downarrow  \right)}-E_{\rm o}^{\uparrow \left( \downarrow  \right)}}}\  \\ \nonumber
+ \underbrace{\frac{{{\left| \left\langle  \Psi _{\rm o}^{\uparrow \left( \downarrow  \right)} \right|{{{\hat{L}}}_{x}}\left| \Psi _{\rm u}^{\downarrow \left( \uparrow  \right)} \right\rangle  \right|}^{2}}-{{\left| \left\langle  \Psi _{\rm o}^{\uparrow \left( \downarrow  \right)} \right|{{{\hat{L}}}_{z}}\left| \Psi _{\rm u}^{\downarrow \left( \uparrow  \right)} \right\rangle  \right|}^{2}}}{E_{\rm u}^{\uparrow \left( \downarrow  \right)}-E_{\rm o}^{\downarrow \left( \uparrow  \right)}}}_{\text{spin-flip}\ \text{term}}
  \label{eqn:graphic}
\end{eqnarray}
where $\xi$ is the SOC constant, $\Psi_{\rm o}^{\uparrow (\downarrow)}$ and $\Psi_{\rm u}^{\uparrow (\downarrow)}$ indicate respectively the occupied and unoccupied majority-spin (minority-spin) bands, and $E_{\rm o}^{\uparrow (\downarrow)}$ and $E_{\rm u}^{\uparrow (\downarrow)}$ respectively represent the corresponding energies. For the same spin, SOC between occupied and unoccupied states with the same (different) magnetic quantum number through the $\hat{L}_z$ ($\hat{L}_x$) operator gives a positive (negative) contribution to MCA. SOC between opposite spin states gives reverse contribution to MCA.

Let us use the H-FeSe$_2$ monolayer as an example to show how the external field modifies the band structure and the perturbation terms in the above expression for MCA. We have made this choice because the effect is most significant in this material (Fig. 3). Due to the crystal field effect in trigonal prisms, $d$-orbitals split into three groups, $a$ ($d_{z^2}$) ($\left| m \right| = 0$), $e_1$ ($d_{yz}, d_{xz}$) ($\left| m \right| = 1$) and $e_2$ ($d_{x^2-y^2}, d_{xy}$) ($\left| m \right| = 2$), which are denoted by blue, green and red colors, respectively. For FeSe$_2$ under zero field [Fig. 3(a)], the coupling between majority spin (left panel) including $e_1$, and including $e_2$, through the $\hat{L}_x$ operator give negative contribution to MCA. This happens both along the ($\Gamma$-K) high symmetry line (labeled as HSL1) as well as along the ($\Gamma$-M) (labeled as HSL2) directions. For the minority spin (in right panel and labeled as HSL3), the coupling between occupied $a$ and unoccupied $e_2$ through the $\hat{L}_x$ operator also gives negative contribution.

In next step, an electric field perpendicular to the TMD plane is applied. It causes Fe $d$-orbitals shift downwards for the majority-spin band, which substantially reduces the proportion of $d$-orbitals around the Fermi level (left panels), especially for the states around HSLn ($n = 1$ and 2). When electric field reaches 0.75 V/{\AA}, $d$ states almost vanish around the Fermi level, except for a very small proportion of $e_1$ states [Fig. 3(d) left panel]; hence the negative contribution to MCA around HSL1 and HSL2 decrease greatly. Considering the spin flip terms, the minority bands of $e_2$ states below the Fermi energy (occupied) shift upwards [HSL4 and HSL5 in the right panels of Fig. 3(a) - (d)], whereas the majority bands of the unoccupied $e_1$ states along the ($\Gamma$-K) (HSL4) and (M-K) (HSL5) shift downwards [Fig. 3(c) and (d) left panels]. As a consequence, the energy differences between the minority occupied $e_{2}$ states and majority unoccupied $e_{1}$ states decrease. This causes stronger coupling between these two opposite spin states through the $\hat{L}_{x}$ operator, which gives a large positive contribution to MCA. Therefore, with the increasing of electric field, the above negative terms contribute less to MCA whereas the positive terms contribute more, which eventually causes MCA change from negative to positive when the field is high enough.

\begin{figure} [ht]
 \centering
 \includegraphics[width=0.35\textwidth]{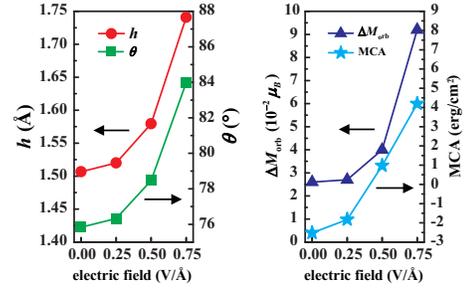}\\
 \caption{(Color online) The variation of relevant structural parameters verse electric field for FeSe$_2$ system. $h$ ({\AA}) and $\theta$ ($^{\circ}$) represent, respectively, the vertical distance between the chalcogen atom and the transition metal atom, and the X-A-X bond angles. While $\Delta M_{\rm orb}$ (in units of $\mu_B$) is the orbital moment anisotropy.}
 \label{fig:Figure4}
\end{figure}

Figure 4 shows how the geometry parameters and orbital moment anisotropy ($\Delta M_{\text{orb}} = M_{\text{orb}[001]} - M_{\text{orb}[100]}$) change under an electric field. It reveals that the geometry changes under an electric field together with the change of orbital momentum and MCA. The vertical distances ($h$) between the chalcogen atom and the transition metal atom change non-linearly from 1.51 {\AA} to 1.74 {\AA}.  By increasing the intensity of the electric field, $h$ grows rapidly. The same behavior also happens to X-A-X bond angles $\theta$, which changes from $75.8^{\circ}$ to $84.0^{\circ}$. Both of the variation trends are parallel with that of the MCA. Despite the applied field breaks the $D_{3h}$ symmetry, all six A-X bond lengths remain the same. More significant changes happen to the orbital moments. $\Delta M_{\text{orb}}$ changes from 0.026 $\mu_{B}$ to 0.092 $\mu_{B}$ as the electric field increases from 0 to 0.75 V/{\AA}. And it is commonly known that higher orbital moments as well as their anisotropy usually indicate larger MCA. The changes in the orbital moment of the AX$_2$ monolayers under electric field are very different from those of thin transition metal films, which are much smaller and often nonmonotonic\cite{surface_metal,Kioussis,linear}. It is worth noticing that similar geometry changes under electric field  can be observed for most of the 2D AX$_2$ monolayers. However, only those with strong SOC and large MCA show significant variations of the orbital momentum as well as the voltage modulation of MCA. Our results suggest that materials with both strong SOC and strong covalent bonds are good choices for voltage-controllable MCA.

Inspired by the above results, that the geometry and magnetic properties including MCA change simultaneously for AX$_2$ monolayers under increasing electric field, we also explored how the biaxial strain could change the MCA values. Our results reveal a strong strain effect (refer to Supporting Information Fig. S3). Among all the AX$_2$ monolayers, T-CrSe$_2$, H-FeSe$_2$ and H-TaS$_2$ show exceedingly large strain effect on MCA. Particularly, the MCA of T-CrSe$_2$ increases nonlinearly with an applied tensile biaxial strain and changes from in-plane to out-of-plane at strain of about 2\%. It is well known that monolayer materials may sustain large biaxial strains. Although, the direct control of MCA by strain is not easy to achieve, the combination of strain and voltage may greatly extend our control of MCA in nano-sized devices.

In summary, we propose that single layer transition metal dichalcogenides can be ideal candidate materials for voltage controlled memory devices. Using first principles DFT calculations, we demonstrate that 2D AX$_2$ materials may exhibit both colossal MCA and exceedingly strong voltage dependence, and their easy-axes change from in-plane at zero field to out-of-plane under finite electric field. The polarized covalent bonds signify the SOC on heavy atoms, causing large MCA. Comparing with thin metal films, these covalently bonded materials exhibit large geometry deformation under electric filed and much smaller screening effect, resulting at an enormous voltage modulation to its MCA. Hence, these materials can meet the two opposing demands for the new type of magnetic memory, namely maintaining the memory against thermodynamic fluctuations and writing or rewriting with low power consumption. This makes them excellent candidates for future memory devices.

We acknowledge the support of the Ministry of Science and Technology of China (Grant No. 2016YFA0301001), and the National Natural Science Foundation of China (Grants No. 11674188 and 11334006). Some calculations are performed on NSF-funded XSEDE resources (TG-DMR130005) especially on Stampede cluster ran by TACC.


\begin{thebibliography}{99}

\bibitem{nature} W. Eerenstein, N. D. Mathur, and J. F. Scott, Nature {\bf 442}, 759 (2006).

\bibitem{mutiferroics} R. N. Ramesh, and A. Spaldin, Nat. Mater. {\bf 6}, 21 (2007).

\bibitem{spin} C. Chappert, A. Fert, and F. N. Van Dau, Nat. Mater. \emph{6}, 813 (2007).

\bibitem{science} M. Weisheit, S. Fhler, A. Marty, Y. Souche, C. Poinsignon, and D. Givord, Science {\bf 315}, 349 (2007).

\bibitem{surface_metal} C.-G. Duan, J. P. Velev, R. F. Sabirianov, Z. Zhu, J. Chu, S. S. Jaswal, and E. Y. Tsymbal, Phys. Rev. Lett. {\bf 101}, 137201 (2008).

\bibitem{monolayer} K. Nakamura, R. Shimabukuro, Y. Fujiwara, T. Akiyama, T. Ito, and A. J. Freeman, Phys. Rev. Lett. {\bf 102}, 187201 (2009).

\bibitem{fewlayeriron} T. Maruyama, Y. Shiota, T. Nozaki, K. Ohta, N. Toda, M. Mizuguchi, A. A. Tulapurkar, T. Shinjo, M. Shiraishi, S. Mizukami, Y. Ando, and Y. Suzuki, Nat. Nanotechnol. {\bf 4}, 158 (2009).

\bibitem{FeCo} Y. Shiota, T. Nozaki, F. Bonell, S. Murakami, T. Shinjo, and Y. Suzuki, Nat. Mater. {\bf 11}, 39 (2012).

\bibitem{Au/FeCo} P. V. Ong, N. Kioussis, P. Khalili Amiri, and K. L. Wang, Sci. Rep. {\bf 6}, 29815 (2016).

\bibitem{junction} Z. Bai, L. Shen, Y. Cai, Q, Wu, M. Zeng, G. Han and Y. P. Feng, New J. Phys. {\bf 16}, 103033 (2014).

\bibitem{junction2} S. Kanai, M. Yamanouchi, S. Ikeda, Y. Nakatani, F. Matsukura, H. Ohno, Appl. Phys. Lett. {\bf 101}, 122403 (2012).

\bibitem{Dimer} J. Hu, and R. Wu, Nano Lett. {\bf 14}, 1853 (2014).

\bibitem{Kioussis} P. V. Ong, N. Kioussis, P. K. Amiri, J. G. Alzate, K. L. Wang, G. P. Carman, J. Hu, and R. Wu, Phys. Rev. B {\bf 89}, 094422 (2014).

\bibitem{geim2007rise} Geim, A. K.; Novoselov, K. S. Nat. Mater. {\bf 6}, 183 (2007).

\bibitem{butler2013progress} S. Z. Butler, S. M. Hollen, L. Cao, Y. Cui, J. A. Gupta, H. R. Guti{\'e}rrez, T. F. Heinz, S. S. Hong, J. Huang, A. F. Ismach, et al. ACS Nano {\bf 7}, 2898 (2013).

\bibitem{two-dimensional} M. Xu, T. Liang, M. Shi, H. Chen, Chem. Rev. {\bf 113}, 3766 (2013).

\bibitem{yan2007} Q. Yan, B. Huang, J. Yu, F. Zheng, J. Zang, J. Wu, B.-L. Gu, F. Liu, and W. Duan, Nano Lett. {\bf 7}, 1469 (2007)

\bibitem{defects} X. Zou, and B. I. Yakobson, Acc. Chem. Res. {\bf 48}, 73 (2015).

\bibitem{synthetic} K.-K. Liu, W. Zhang, Y.-H. Lee, Y.-C. Lin, M.-T. Chang, C.-Y. Su, C.-S. Chang, H. Li, Y. Shi, H. Zhang, C.-S. Lai, and L.-J. Li, Nano Lett. {\bf 12}, 1538 (2012).

\bibitem{exfoliation} J. N. Coleman, M. Lotya, A. O'Neill, S. D. Bergin, P. J. King, U. Khan, K. Young, A. Gaucher, S. De, R. J. Smith, et al. Science {\bf 331}, 568 (2011).

\bibitem{single_layer} Z. Zeng, Z. Yin, X. Huang, H. Li, Q. He, G. Lu, F. Boey, and H. Zhang, Angew. Chem. Int. Ed. {\bf 50}, 11093 (2011).

\bibitem{mechanical} H. Li, G. Lu, Y. Wang, Z. Yin, C. Cong, Q. He, L. Wang, F. Ding, T. Yu, and H. Zhang, Small, {\bf 9}, 1974 (2013).

\bibitem{stable} C. Ataca, H. Sahin, and S. Ciraci, J. Phys. Chem. C {\bf 116}, 8983 (2012).

\bibitem{VS2} Y. Ma, Y. Dai, M. Guo, C. Niu, Y. Zhu, B. Huang, ACS Nano {\bf 6}, 1695 (2012).

\bibitem{grain} Z. Zhang, X. Zou, V. H. Crespi, B. I. Yakobson, ACS Nano {\bf 7}, 10475 (2013).

\bibitem{magnetic} H. Guo, N. Lu, L. Wang, X. Wu, and X. C. Zeng, J. Phys. Chem. C {\bf 118}, 7242 (2014).

\bibitem{Fe/MoS} W. T. Cong, Z. Tang and J. H. Chu, Sci. Rep. {\bf 5}, 9361 (2015).

\bibitem{Fe/TaS} V. Loganathan, J.-X. Zhu and A. H. Nevidomskyy, arXiv: 1605.07141 (2016).

\bibitem{DFT} W. Kohn, and L. J. Sham, Phys. Rev. {\bf 140}, A1133(1965).

\bibitem{VASP} G. Kresse, and J. Furthm{\"u}ller, Comput. Mater. Sci. {\bf 6}, 15 (1996).

\bibitem{PAW} Bl{\"o}chl, P. E. Phys. Rev. B {\bf 50}, 17953 (1994).

\bibitem{PBE} Perdew, J. P.; Burke, K.; Ernzerhof, M. Phys. Rev. Lett. {\bf 77}, 3865 (1996).

\bibitem{LDA+U} V. I. Anisimov, F. Aryastiawan, A. I. Lichtenstein, J. Phys.: Condens. Matter {\bf 9}, 767 (1997).

\bibitem{HSE} J. Heyd, G. E. Scuseria, M. Ernzerhof, J. Chem. Phys. {\bf 118}, 8207 (2003).

\bibitem{thin_TaS2} E. Navarro-Moratalla, J. O. Island, S. Ma{\~n}as-Valero, E. Pinilla-Cienfuegos, A. Castellanos-Gomez, J. Quereda, G. Rubio-Bollinger, L. Chirolli, J. A. Silva-Guill{\'e}n, N. Agra{\"i}t, et al. Nat. Commun. {\bf 7}, 11043(2016).

\bibitem{MCA} D. Wang, R. Wu, and A. J. Freeman, Phys. Rev. B {\bf 47}, 14932 (1993).

\bibitem{origin} P. Bruno, Physical Origins and Theoretical Models of Magnetic Anisotropy (Ferienkurse des Forschungszentrums J{\"u}lich, J{\"u}lich, Chapter: 24) (1993).

\bibitem{linear} M. Tsujikawa, and T. Oda, Phys. Rev. Lett. {\bf 102}, 247203 (2009).


\end{thebibliography}
\end{document}